\documentclass[pra,aps,twocolumn]{revtex4}
\usepackage{graphicx}
\usepackage{amsmath}

\begin{document}
\title{Improving fidelity in atomic state teleportation via cavity
  decay}

\author{Grzegorz Chimczak and Ryszard Tana\'s}

\affiliation{Nonlinear Optics Division, Physics Institute, Adam
  Mickiewicz University, 61-614 Pozna\'n, Poland}

\date{\today} \email{chimczak@kielich.amu.edu.pl}

\begin{abstract}
  We propose a modified protocol of atomic state teleportation for the
  scheme proposed by Bose~\emph{et al.} (\emph{Phys. Rev. Lett.}
  {\textbf{83}}, 5158 (1999)). The modified protocol involves an
  additional stage in which quantum information distorted during the
  first stage is fully recovered by a compensation of the damping
  factor.  The modification makes it possible to obtain a high
  fidelity of teleported state for cavities that are much worse than
  that required in the original protocol, i.e., their decay rates can
  be over $25$ times larger. The improvement in the fidelity is
  possible at the expense of lowering the probability of success.  We
  show that the modified protocol is robust against dark counts.
\end{abstract}

\maketitle

\thispagestyle{empty}

\section{Introduction}
Quantum teleportation~\cite{bennett_tele} is considered to be a
perfect way of transferring qubits over long distances. It is
particularly important to teleport qubits represented by the atomic
states, which can store quantum information for sufficiently long time
as to make it available for further quantum processing. However, in
contrast to the teleportation of photonic states, the teleportation of
atomic states over long distances is a difficult task. As yet, the
longest distance achieved experimentally for atomic states is of the
order of micrometers~\cite{barrett04,riebe04} while for photonic
states is of the order of kilometers~\cite{marcikic03}. It is obvious
that the distance of atomic states teleportation has to be orders of
magnitude greater to make the teleportation useful in quantum
communication.  In order to make this distance greater, it is
necessary to employ photons, which are the best long distance carriers
of quantum information, to establish quantum communication between the
remote atoms and complete the atomic state teleportation. Such a
scheme of atomic state teleportation has been presented by
Bose~\emph{et al.}~\cite{bose}.  They have proposed an additional
stage of teleportation protocol --- the preparation stage, in which
the state of sender atom is mapped onto the sender cavity field state
and therefore can be teleported in the next stage using well known
linear optics techniques.  The possibility of operating on atomic
qubits with linear optics elements is the reason why a combination of
atomic states and cavity field states has been recently suggested in many
proposals, not only in proposals of teleportation
protocols~\cite{chimczak:_entanglement_teleportation,xue06:_telep_qed,zheng06:_schem}
but also in other schemes of quantum information
processing~\cite{chimczak:_entanglement,lim05:_repeat_lett,lim06:_repeat, 
  cao06:_concen,guo06:_effec,metz:_robustEntanglement,beige:_multi}.
Unfortunately, the state mapping and whole preparation stage is not
perfect because of a destructive role played by cavity decay.  The
cavity decay reduces the fidelity of teleported state and the
probability of success. Bose~\emph{et al.}~\cite{bose} have suggested
a way to minimize a destructive role of this imperfection by assuming
very small cavity decay rate. However, the value of cavity decay
rate required by their protocol is two orders of magnitude below of
what is currently available~\cite{hennrich00,
  kuhn02,mckeever03:_state,mckeever03:_exper,mckeever04:_single_photon,
  legero04,boozer06:_cooling,aoki06:_observation}.

Here, we present a protocol that reduces the effect of
cavity decay on the fidelity. This protocol makes it possible to use
cavities with larger decay rates without worsening the fidelity but at
the expense of lowering success rates.

\section{Model}
\begin{figure}[htbp]
  \centering
  \includegraphics[width=7cm]{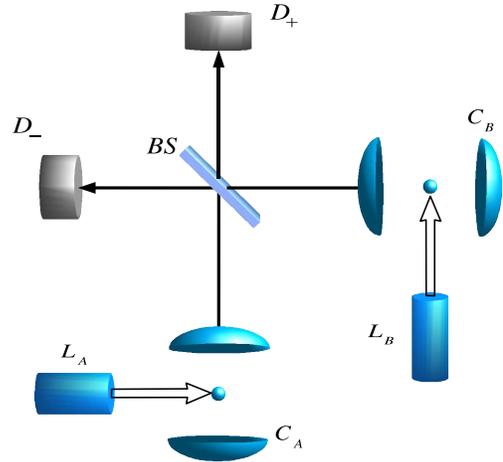}
  \caption{(Color online) Schematic representation of the setup to
    realize long distance teleportation of atomic states via photons.}
  \label{fig:device}
\end{figure}

\begin{figure}[htbp]
  \centering
  \includegraphics[width=7cm]{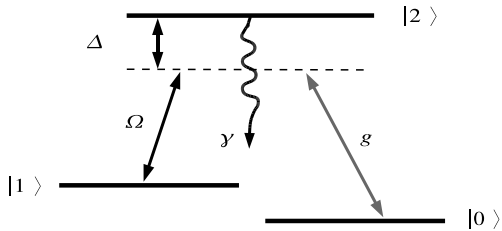}
  \caption{Level scheme of the $\Lambda$ atom interacting with the
    classical laser field and the quantized cavity mode.}
  \label{fig:lambda}
\end{figure}

The teleportation protocol that we propose in this paper is designed
for the same device which Bose \emph{et al.}~\cite{bose} consider in their scheme.
The device is depicted in Fig.~\ref{fig:device}. It is
composed of two cavities $C_{A}$ and $C_{B}$, a 50-50 beam splitter,
two lasers $L_{A}$ and $L_{B}$ and two single-photon detectors $D_{+}$
and $D_{-}$. The receiver, Bob, has the cavity $C_{B}$ and the laser
$L_{B}$. The other elements of the device are at the side of the
sender --- Alice. Inside each cavity there is one trapped atom,
modeled by a three-level $\Lambda$ system with two stable ground
states $|0\rangle$ and $|1\rangle$, and one excited state $|2\rangle$
as shown in Fig.~\ref{fig:lambda}.
Only the excited state decays spontaneously, therefore the ground
states are ideal candidates for an atomic qubit. The spontaneous decay
rate of the excited state is denoted by $\gamma$. Operations on the
qubit coded in the superposition of both ground states are possible
using two transitions: $|0\rangle \leftrightarrow |2\rangle$ and
$|1\rangle \leftrightarrow |2\rangle$. First of the transitions is
coupled to the cavity mode with the coupling strength $g$ while the
second transition is coupled to a classical laser field with the
coupling strength $\Omega$.  Since we want a population of the excited
state to be negligible, the laser field and the cavity mode are
detuned from the corresponding transition frequencies by $\Delta$.
Beside the spontaneous decay of the excited atomic state there is
another decay mechanism. One mirror in each cavity is partially
transparent and therefore photons leak out of the cavities through
these mirrors at a rate $\kappa$. The evolution of each atom-cavity
system is governed by the effective non-Hermitian Hamiltonian
($\hbar=1$ here and in the following)
\begin{eqnarray}
  \label{eq:Hamiltonian0}
  H&=&(\Delta - i \gamma) \sigma_{22} 
  +(\Omega \sigma_{21}+g a \sigma_{20}+ {\rm{H.c.}})
  -i \kappa a^{\dagger} a \, , \nonumber \\
\end{eqnarray}
where $a$ denotes the annihilation operator for Alice's cavity mode
($a_{A}$) or Bob's cavity mode ($a_{B}$).  In~(\ref{eq:Hamiltonian0})
we introduce the flip operators $\sigma_{ij}\equiv |i\rangle \langle
j|$.  Both our protocol as well as the Bose~\emph{et al.}  protocol
work in the low saturation limit ($g^2/\Delta^2$,
$\Omega^2/\Delta^2\ll 1$) and therefore the excited atomic state can
be adiabatically
eliminated~\cite{buck_adiaba,singh_adiaba,knight_adiaba,puri_adiaba,
  phoenix_adiaba,carmichaelKsiazkaMethods,pell,alexanian95}.  Either
of them require small values of the spontaneous decay rate
($\Delta\gg\gamma$ and $\gamma g^2/\Delta^2$,
$\gamma\Omega^2/\Delta^2\ll\kappa$)~\cite{chimczak02:_effect} which
makes it possible to neglect $\gamma$ as a first approximation.
Finally, we assume the condition $\Omega=g$ which leads to a very
simple form of the Hamiltonian
\begin{eqnarray}
  \label{eq:Hamil1}
  H=-\delta \sigma_{11}-\delta a^{\dagger} a \sigma_{00}
  -(\delta a \sigma_{10} +{\rm{H.c.}})
  -i \kappa a^{\dagger} a \, ,
\end{eqnarray}
where $\delta=g^2/\Delta$.  The evolution described
by~(\ref{eq:Hamil1}) is interrupted by collapses.  Photon decays
registered by detectors correspond to the action of the collapse
operator
\begin{eqnarray}
  \label{eq:C}
  C=\sqrt{\kappa}(a_{A}+i\epsilon a_{B}) \, ,
\end{eqnarray}
where $\epsilon$ is equal to $1$ for photon detection in $D_{+}$ and
equal to $-1$ for photon detection in $D_{-}$.

The simple form of the Hamiltonian~(\ref{eq:Hamil1}) allows for
analytical solutions of the non-unitary Schr\"odinger equation and get
expressions for the time evolution of quantum states which are used in
both protocols.  To give the expressions a more compact form
we use the notation $|j n\rangle\equiv
|j\rangle_{\textrm{atom}}\otimes|n\rangle_{\textrm{mode}}$ to describe
the state of the atom-cavity system.  During the whole teleportation
process the time evolution of the system is restricted to the subspace
spanned by the states: $|00\rangle$, $|10\rangle$ and $|01\rangle$.
The state $|00\rangle$ experiences no dynamics because there is no
operator in the Hamiltonian~(\ref{eq:Hamil1}) which can change this
state. Time evolution of the other two states is described by
\begin{eqnarray}
  \label{eq:U}
  e^{-i H t} |10\rangle&=& 
  e^{i \delta t} e^{-\frac{\kappa t}{2}} \Big[
  i \frac{2 \delta}{\Omega_{\kappa}} 
  \sin\Big(\frac{\Omega_{\kappa} t}{2}\Big) |01\rangle \nonumber \\
  &&+\Big(\cos\Big(\frac{\Omega_{\kappa} t}{2}\Big)
  +\frac{\kappa}{\Omega_{\kappa}} 
  \sin\Big(\frac{\Omega_{\kappa} t}{2}\Big)\Big) |10\rangle 
  \Big] \, , \nonumber \\ 
  e^{-i H t} |01\rangle&=&
  e^{i \delta t} e^{-\frac{\kappa t}{2}} \Big[
  i \frac{2 \delta}{\Omega_{\kappa}} 
  \sin\Big(\frac{\Omega_{\kappa} t}{2}\Big) |10\rangle \nonumber \\
  &&+\Big(\cos\Big(\frac{\Omega_{\kappa} t}{2}\Big)
  -\frac{\kappa}{\Omega_{\kappa}} 
  \sin\Big(\frac{\Omega_{\kappa} t}{2}\Big)\Big) |01\rangle 
  \Big] \, , \nonumber \\
\end{eqnarray}
where $\Omega_{\kappa}=\sqrt{4 \delta^2-\kappa^2}$.  There are two
important local operations we can perform on the system state via
$e^{-i H t}$. First of them is to map the atomic state onto the cavity
mode and second is the generation of the maximally entangled state of
the atom and the cavity mode. The atomic state mapping one can obtain
by turning the laser on for time $t_{A}$ is given by
\begin{eqnarray}
  \label{eq:U1}
  |10\rangle &\rightarrow &
  i e^{i \delta t_{A}} e^{-\frac{\kappa t_{A}}{2}} |01\rangle \, ,
\end{eqnarray}
where
$t_{A}=(2/\Omega_{\kappa})[\pi-\arctan(\Omega_{\kappa}/\kappa)]$.  In
order to create the maximally entangled state the laser should be
turned on for time
$t_{B}=(2/\Omega_{\kappa})\arctan(\Omega_{\kappa}/(2\delta-\kappa))$
\begin{eqnarray}
  \label{eq:OS1}
  |10\rangle&\rightarrow& e^{i \delta t_{B}} e^{-\frac{\kappa t_{B}}{2}}  
  \frac{2 \delta}{\Omega_{\kappa}} 
  \sin\Big(\frac{\Omega_{\kappa} t_{B}}{2}\Big)
  ( |10\rangle +i |01\rangle ) \, .
\end{eqnarray}

When the laser is turned off then $\Omega=0$, and the Hamiltonian goes
over into $H=-\delta a^{\dagger} a \sigma_{00}-i \kappa a^{\dagger}
a$.  Then all the terms of the Hamiltonian correspond to the diagonal
elements in matrix representation, and the non-unitary Schr\"odinger
equation can be easily solved.  The evolution of the states
$|10\rangle$ and $|01\rangle$, when the laser is turned off, are thus
given by
\begin{eqnarray}
  \label{eq:las0a}
  e^{-i H t} |10\rangle&=& |10\rangle \, , \nonumber \\
  \label{eq:las0b}
  e^{-i H t} |01\rangle&=& e^{i \delta t} e^{-\kappa t}|01\rangle  \, .
\end{eqnarray}

\section{Teleportation protocol}
Both protocols start with the same initial state --- the unknown state
that Alice wants to teleport, which is stored in her atom. Bob's atom
is prepared in the state $|1\rangle$ and the field modes of both
cavities are empty, so we have
\begin{eqnarray}
  \label{eq:p0}
  |\psi\rangle_{A}&=&\alpha|00\rangle_{A} +\beta |10\rangle_{A} \, , \\
  \label{eq:p1}
  |\psi\rangle_{B}&=&|10\rangle_{B} \, .
\end{eqnarray}
The teleportation protocol with improved fidelity consists of five
stages: (A) the preparation stage, (B) the detection stage I, (C) the
compensation stage, (D) the detection stage II and (E) the recovery
stage.

\subsection{Preparation stage}
The preparation stage is necessary because the quantum information
encoded initially in Alice's atom is teleported by performing joint
measurement on the field state of both cavities. Before of the
detection stage Alice has to map the quantum information onto her
cavity field state while Bob has to create the maximally entangled
state of his atom and his cavity field.  Alice and Bob achieve their
goals by switching their lasers on for times $t_{A}$ and $t_{B}$,
respectively.  After the preparation stage the state of Alice's
atom-cavity system is given by
\begin{eqnarray}
  \label{eq:p3}
  |\widetilde{\psi}\rangle_{A}=\alpha|00\rangle_{A} +i e^{i\delta t_{A}} e^{-\frac{\kappa t_{A}}{2}}\beta |01\rangle_{A} \, ,
\end{eqnarray}
and Bob's system state becomes
\begin{eqnarray}
  \label{eq:p4}
  |\widetilde{\psi}\rangle_{B}=e^{-\frac{\kappa t_{B}}{2}}
  \frac{2 \delta}{\Omega_{\kappa}} 
  \sin\Big(\frac{\Omega_{\kappa} t_{B}}{2}\Big)
  (|10\rangle_{B}+i |01\rangle_{B}) \, .
\end{eqnarray}
This first stage is successful only under the absence of photon
detection event.  The probability that no collapse occurs during
Alice's operation is given by the squared norm of the state vector
\begin{eqnarray}
  \label{eq:Pa}
  P_{A}=|\alpha|^{2} + e^{-\kappa t_{A}} |\beta|^{2} \, .
\end{eqnarray}
Similarly, we can obtain appropriate expression for the probability of
no collapse during Bob's operation
\begin{eqnarray}
  \label{eq:Pb}
  P_{B}=e^{-\kappa t_{B}} \frac{8 \delta^2}{\Omega_{\kappa}^2} \sin^2\Big(\frac{\Omega_{\kappa} t_{B}}{2}\Big) \, .
\end{eqnarray}
It is evident that the state mapping is not perfect because of the
damping factors that appear in expression~(\ref{eq:Pa}) for $P_{A}$
and in expression~(\ref{eq:p3}) for the state
$|\widetilde{\psi}\rangle_{A}$. These damping factors reduce both the
probability that the state mapping is successful and the fidelity of
this operation.  The quantum information after the mapping operation
is also modified by the phase factor $i e^{i\delta t_{A}}$ but, in
contrast to damping factors, the phase factors can later be easily
compensated for and therefore they do not reduce the fidelity.  In
order to make the probability $P_{A}$ and the fidelity close to unity
Bose~\emph{et al.} assume that $\Omega_{\kappa}\gg\kappa$, which means
that both $\kappa$ and $t_{A}$ values are small and the damping factor
$e^{-\kappa t_{A}/2}$ is close to unity. Generally, however, the
damping factor is not unity even for very small $\kappa$ and $t_{A}$
and, in consequence, the fidelity of the teleported state is
diminished. Since high fidelities are required by quantum computation
algorithms, we will show how to compensate for this factor in the next
stages of the protocol.

\subsection{Detection stage I}
When the quantum information is mapped onto the state of Alice's
cavity field and the maximally entangled state of Bob's cavity field
and the target atom is created, then the joint measurement of both
cavity fields can be performed.  During this stage Alice and Bob
perform the joint measurement just by waiting with their lasers turned
off.  The teleportation is successful if the detectors register one
and only one photon. In successful cases the joint state of Alice's
and Bob's systems becomes
\begin{eqnarray}
  \label{eq:tel16} 
  |\widetilde{\phi}(t_{d})\rangle&=&
  (i \epsilon \alpha |00\rangle_{B} 
  +e^{i \delta t_{A}} e^{-\frac{\kappa t_{A}}{2}} \beta |10\rangle_{B}) 
  |00\rangle_{A} \nonumber \\
  &&+i e^{i \delta t_{1}}  e^{-\frac{\kappa t_{1}}{2}} \beta e^{-\kappa t_{d}} 
  e^{i \delta t_{d}} \nonumber \\
  && \times (|01\rangle_{B} |00\rangle_{A}+i \epsilon |00\rangle_{B} |01\rangle_{A}) \, ,
\end{eqnarray}
where $t_{d}$ is the time of this detection stage. Until now the
operations in both protocols are exactly the same. In the protocol of
Bose \emph{et al.} it is assumed that time $t_{d}$ is much longer than
$\kappa^{-1}$ and thus all unwanted states in
expression~(\ref{eq:tel16}) can be neglected.  Finally, after removing
a phase factor, the state of Bob's atom is given by $\alpha
|0\rangle_{B}+e^{-\kappa t_{A}/2} \beta |1\rangle_{B}$.  It is obvious
that the fidelity of teleported state will never reach unity because
of the factor $e^{-\kappa t_{A}/2}$.  Moreover, in the protocol of
Bose~\emph{et al.} the fidelity of teleported state decreases with
increasing $\kappa$. In our protocol, we use one of the unwanted
states to compensate for the factor $e^{-\kappa t_{A}/2}$.  This compensation
can be done if we choose the time of this detection stage such that
$e^{i \delta t_{d}}=-1$. Then expression~(\ref{eq:tel16}) can be
rewritten as
\begin{eqnarray}
  \label{eq:tel32} 
  |\widetilde{\phi}(t_{d})\rangle&=&
  i \epsilon \alpha |00\rangle_{B} |00\rangle_{A}
  + e^{i \delta t_{A}}  e^{-\frac{\kappa t_{A}}{2}} 
  \beta e^{-\kappa t_{d}} 
  \epsilon |00\rangle_{B} |01\rangle_{A} \nonumber \\
  &&+e^{i \delta t_{A}}  e^{-\frac{\kappa t_{A}}{2}} \beta 
  (|10\rangle_{B}-i e^{-\kappa t_{d}} |01\rangle_{B}) |00\rangle_{A}  \, .
\end{eqnarray}
\subsection{Compensation stage}
In the compensation stage Bob compensates for the factor $e^{-\kappa t_{A}/2}$ by
turning his laser on for time $t_{c}$. During the operation Alice's
laser remains turned off. On condition that no photon detection occurs
during time $t_{c}$, the unnormalized joint state at the end of this
stage is given by
\begin{eqnarray}
  \label{eq:tel32b} 
  |\widetilde{\phi}(t_{c})\rangle&=&
  e^{i \delta (t_{A}+t_{c})}  e^{-\frac{\kappa t_{A}}{2}} 
  \beta e^{-\kappa (t_{d}+t_{c})} \epsilon 
  |00\rangle_{B} |01\rangle_{A} \nonumber \\
  &&-i e^{i \delta (t_{A}+t_{c})} \beta e^{-\frac{\kappa (t_{A}+t_{c})}{2}} \varphi(t_{c})  
  |01\rangle_{B} |00\rangle_{A} \nonumber \\
  &&+e^{i \delta (t_{A}+t_{c})} \beta e^{-\frac{\kappa (t_{A}+t_{c})}{2}} 
  \vartheta(t_{c}) |10\rangle_{B} |00\rangle_{A}  \nonumber \\
  &&+i \epsilon \alpha |00\rangle_{B} |00\rangle_{A} \, ,
\end{eqnarray}
where
\begin{eqnarray}
  \label{eq:tel33} 
  \varphi(t_{c})&=&
  e^{-\kappa t_{d}}\cos\Big(\frac{\Omega_{\kappa} t_{c}}{2}\Big)
  -\frac{2 \delta+\kappa e^{-\kappa t_{d}}}{\Omega_{\kappa}} 
  \sin\Big(\frac{\Omega_{\kappa} t_{c}}{2}\Big) \, ,
  \nonumber \\
  \vartheta(t_{c})&=&
  \cos\Big(\frac{\Omega_{\kappa} t_{c}}{2}\Big)
  +\frac{\kappa+2 \delta e^{-\kappa t_{d}}}{\Omega_{\kappa}} 
  \sin\Big(\frac{\Omega_{\kappa} t_{c}}{2}\Big) \, .
\end{eqnarray}
It is seen that this operation transfers population from the state
$|01\rangle_{B} |00\rangle_{A}$, which is unwanted, to the state
$|10\rangle_{B} |00\rangle_{A}$. Of course, we want the transfer to
compensate for the factor $e^{-\kappa t_{A}/2}$ and therefore $t_{c}$
has to fulfill the condition
\begin{eqnarray}
  \label{eq:co} 
  e^{-\frac{\kappa (t_{A}+t_{c})}{2}} \vartheta(t_{c})&=&1 \, .
\end{eqnarray}

\subsection{Detection stage II}
The population of one of the two unwanted states is already reduced
after the previous stage, but it cannot be neglected yet. Moreover, the
population of the second unwanted state is still considerable.
Presence of the two unwanted states decreases the teleportation fidelity, so in
the fourth stage of the protocol Alice and Bob have to eliminate them.
All they have to do to achieve this goal is simply to wait
for a finite time $t_{D}\gg \kappa^{-1}$.  After time $t_{D}$ the
populations of both unwanted states are negligible and unnormalized
joint state can be very well approximated by
\begin{eqnarray}
  \label{eq:tel34} 
  |\widetilde{\phi}(t_{D})\rangle&=&
  (i \epsilon \alpha |00\rangle_{B}
  +e^{i \delta (t_{A}+t_{c})} \beta |10\rangle_{B}) |00\rangle_{A} \, .
\end{eqnarray}
\subsection{Recovery stage}
Finally, Bob has to remove the phase shift factor $i \epsilon e^{-i
  \delta (t_{A}+t_{c})}$ to recover the original Alice's state.  To
this end Bob adds to the state $|1\rangle_{\textrm{atom} B}$ an extra
phase shift with respect to the state $|0\rangle_{\textrm{atom} B}$
using the Zeeman evolution~\cite{bose}. After this operation the state
of Bob's atom is exactly the same as the initial state of Alice's atom,
i.e., $\alpha |0\rangle_{\textrm{atom} B}+\beta
|1\rangle_{\textrm{atom} B}$, and thus the teleportation fidelity of
this protocol can be very close to unity.  This completes
the teleportation protocol.

Now it is time to explain in detail how to choose the time $t_{d}$.
The condition $e^{i \delta t_{d}}=-1$ leads to many solutions given by
$t_{d}=\pi (2 m+1)/\delta$, where $m$ is a nonnegative integer.
However, we cannot set $m$ arbitrary because $\vartheta(t_{c})$ and
the probability of success in second stage are functions of $t_{d}$.
It is obvious that the probability of observing one photon during
detection time $t_{d}$ increases with increasing $t_{d}$.  On the
other hand, we cannot choose this detection time too long because the
population of unwanted state $|01\rangle_{B} |00\rangle_{A}$ can then
be too small to compensate for the factor $e^{-\kappa t_{A}/2}$.
Thus, $t_{d}$ is limited by some time $t_{d\textrm{max}}$.  Let us now
estimate $t_{d\textrm{max}}$. Expression $e^{-\kappa (t_{A}+t_{c})/2}
\vartheta(t_{c})$ takes its maximal value for the time of the
compensation stage given by
\begin{eqnarray}
  \label{eq:trmax}
  t_{c \textrm{max}}&=&\frac{2}{\Omega_{\kappa}} 
  \arctan\Bigg(\frac{2\delta\Omega_{\kappa} e^{-\kappa t_{d}}}
  {\Omega_{\kappa}^2+\kappa (\kappa+2\delta e^{-\kappa t_{d}})}\Bigg) \, .
\end{eqnarray}
The factor $e^{-\kappa t_{A}/2}$ can be compensated for only under the
condition that $e^{-\kappa (t_{A}+t_{c \textrm{max}})/2}
\vartheta(t_{c \textrm{max}})\geq 1$.  Since both $\vartheta(t_{c})$
  and $t_{c\text{max}}$ depend on $t_{d}$, we can estimate the
  value of 
$t_{d\textrm{max}}$ by finding numerically $t_{d}$ satisfying the
condition
\begin{eqnarray}
  \label{eq:tdmax}
  e^{-\kappa (t_{A}+t_{c \textrm{max}})/2} \vartheta(t_{c \textrm{max}})=1 \, .
\end{eqnarray}

The problem of choosing $t_{d}$ is much simpler when we want to
compensate for the factor $e^{-\kappa t_{A}/2}$ for as large $\kappa$
as possible.
\begin{figure}[htbp]
  \centering
  \includegraphics[width=7cm]{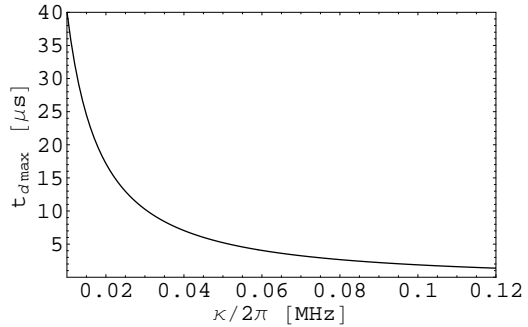}
  \caption{The value of $t_{d\textrm{max}}$ as a function of $\kappa$ for
    $(\Delta;\Omega;g)/2\pi=(100;10;10)$ MHz calculated numerically
    using condition~(\ref{eq:tdmax}).}
  \label{fig:tdodk}
\end{figure}
From figure~\ref{fig:tdodk} one can see that the limit
$t_{d\textrm{max}}$ decreases with increasing $\kappa$.  Therefore we
should choose the smallest value of $t_{d}$ by setting $m$ to zero.

\section{Numerical results}
Let us now compare both protocols. For this purpose we compute the
average probability of success and the average fidelity of teleported
state for the same values of the detuning and both coupling strengths
as in Ref.~\cite{bose}, i.e., $(\Delta;\Omega;g)/2\pi=(100;10;10)$
MHz.  It is necessary to take average values over all input states
because the probability of success in both protocols as well as the
fidelity in the Bose~\emph{et al.} protocol all depend on the unknown
moduli of the amplitudes $\alpha$ and $\beta$ of the initial state.
The fidelity in our protocol seems to be independent of the amplitudes
of initial state and should be equal to unity.
However, this is only true for the simplified model for which the
excited state is eliminated. In more general model described by the
Hamiltonian~(\ref{eq:Hamiltonian0}) the population of the excited
state has a nonzero value during the evolution given by~(\ref{eq:U})
even if the atom is initially prepared in its ground state. However,
the population of the excited state remains zero for the initial state
$|00\rangle$ of atom-cavity system because the state experiences no
dynamics. If the initial state is a superposition given
by~(\ref{eq:p0}) then the population of the excited state depends on
the moduli of the amplitudes $\alpha$ and $\beta$.  Since the
population of the excited state reduces the fidelity, it is also
necessary to average the fidelity in our protocol over all input
states.

We compute all the averages numerically using the method of quantum
trajectories~\cite{carmichaelksiazka,pleniotrajek} together with the
Monte Carlo technique. Each trajectory starts with a random initial
state and evolves according to a chosen teleportation protocol. If
measurement indicates success then we calculate the fidelity of
teleported state at the end of the protocol. Otherwise, we reject a
trajectory as unsuccessful. After generating 20 000 trajectories we
average the fidelity over all trajectories and calculate the average
probability of success as a ratio of the number of successful
trajectories to the number of all trajectories.

There are some problems that appear when we use the
Hamiltonian~(\ref{eq:Hamiltonian0}) to simulate performance of our
protocol. First, the fidelity is sensitive to the inaccuracy in
calculations of phase shift factors.  The compensation of the factor
$e^{-\kappa t_{A}/2}$ requires the phase shift of the state
$|01\rangle_{B} |00\rangle_{A}$ relative to the state $|10\rangle_{B}
|00\rangle_{A}$ to be equal to $-i$ as shown in~(\ref{eq:tel32}).
Therefore the time $t_{d}$ of the detection stage I has to satisfy the
condition $e^{i \delta t_{d}}=-1$.  However, the analytical expression
for $\delta$ is derived from the Hamiltonian~(\ref{eq:Hamil1}) and
thus $\exp(i\delta t_{d})$ is only an approximation to the real phase
shift factor. Unfortunately, the population transfer that takes place
in the compensation stage leads to an unknown extra phase shift in the
final state~(\ref{eq:tel34}) when the phase shift between states
$|01\rangle_{B} |00\rangle_{A}$ and $|10\rangle_{B} |00\rangle_{A}$
differs from the expected value $-i$. Of course, this unknown phase
shift cannot be compensated for in the recovery stage, which means
that the fidelity in our protocol can even be smaller than the
fidelity in the Bose~\emph{et al.} protocol.  To overcome this problem
we use a numerical optimization procedure which finds, for the more
general model, such $t_{d}$ that the joint state of Alice's and Bob's
systems becomes as close to the expected state given
by~(\ref{eq:tel32}) as possible.

\begin{figure}[htbp]
  \centering
  \includegraphics[width=7cm]{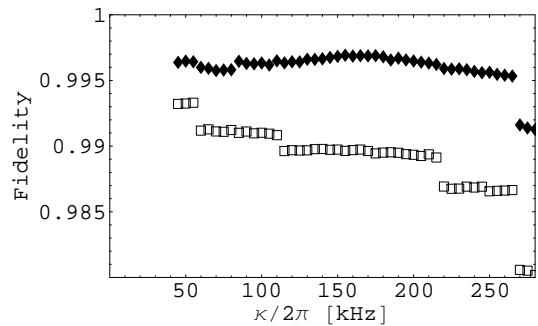}
  \caption{The average fidelity of teleportation in the new protocol
    (diamonds) and in Bose~\emph{et al}. protocol (open squares) as
    functions of the cavity decay rate for
    $(\Delta;\Omega;g;\gamma)/2\pi=(100;10;10;0)$ MHz. The averages
    are taken over 20 000 trajectories.}
  \label{fig:fodk}
\end{figure}

Second, a question arises: how to estimate the biggest value of
$\kappa$ for which the compensation is still possible? This value is
very important because we want to know how good (or rather bad)
cavities can be used
for effective high fidelity teleportation. In the simplified model of
our protocol governed by the Hamiltonian~(\ref{eq:Hamil1}) this
value can be computed from~(\ref{eq:tdmax}) and is about
$\kappa/2\pi\approx 0.17$ MHz.  However, the population of the excited
atomic state changes this value because of the transfer of population
from the state $|20\rangle_{B}|00\rangle_{A}$ to the state
$|10\rangle_{B} |00\rangle_{A}$ in the compensation stage.  To
estimate the acceptable value of $\kappa$, we plot the average
fidelity and the average probability of success as functions of
$\kappa$.  The population of the excited atomic state changes also the
time $t_{c}$ for which the improvement of the fidelity in our protocol
is the best one and thus the value of $t_{c}$ calculated
from~(\ref{eq:co}) can be used only as a starting point in the
numerical computation of this time. From numerical results presented
in Fig.~\ref{fig:fodk} we find that there
is a plateau in the fidelity of the modified protocol up to
$\kappa/2\pi\approx 0.25$ MHz after which the fidelity jumps down. We
consider the value of $\kappa$ at the jump as the biggest value of
$\kappa$.

Third, the population of the excited atomic state oscillates. Since
the population of the excited state diminishes the fidelity of
operations periodically, it is necessary to compute numerically, for
all operations, such times that minimize the population simultaneously
maximizing the fidelity. Until now we have assumed that times $t_{A}$
and $t_{B}$ can be calculated analytically as in the Bose~\emph{et
  al.}  protocol. However, the analytical expressions are
functions of $\kappa$, so, for different values of $\kappa$ the
population of the excited state and the fidelities of operations take
different values. If we want to stabilize the average fidelity at a
high level for different values of $\kappa$ then we have to compute
$t_{A}$ and $t_{B}$ numerically.

To begin with our calculations, we set the spontaneous decay rate of
excited state to zero because we want to know how close to unity is
the fidelity in the ideal case in which there is no possibility of
photon emission to modes other than the cavity modes.
Fig.~\ref{fig:fodk} shows that the modified protocol really stabilizes
the fidelity of teleported state at a high level. The fidelity is
reduced only by the nonzero population of the excited state and does
not decrease with increasing $\kappa$ until $\kappa/2\pi$ is about
$0.25$ MHz.  The fidelity of teleported state in the protocol of
Bose~\emph{et al.} is reduced by the population of excited state as
well as by the factor $\exp(-\kappa t_{A}/2)$ and, as expected, it
decreases with increasing $\kappa$. It is seen from
Fig.~\ref{fig:fodk} that there are discontinuous jumps of the fidelity
values. The discontinuities come from the numerical procedure finding
such $t_{A}$ for which the mapping fidelity is maximal. The time
$t_{A}$ of the mapping operation is a function of $\kappa$, and the
mapping fidelity reaches its maximal value when the population of the
excited state reaches its minimal value. Since the population of the
excited state oscillates the numerically calculated $t_{A}$ jumps, as
$\kappa$ increases, from one value for which the population of the
excited state is minimal after the mapping operation to the next such
value. Thus, the factor $\exp(-\kappa t_{A}/2)$ and the fidelity of
the teleported state also exhibit discontinuous behavior.
\begin{figure}[htbp]
  \centering
  \includegraphics[width=7cm]{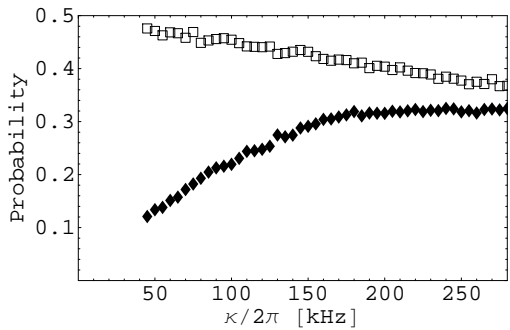}
  \caption{The average probability of successful teleportation as a
    function of the cavity decay rate. The diamonds show the average
    probability of success in the new protocol. The open squares
    correspond to the average probability in Bose~\emph{et al}.
    protocol. The averages are taken over 20 000 trajectories. The
    parameters regime is $(\Delta;\Omega;g;\gamma)/2\pi=(100;10;10;0)$
    MHz.}
  \label{fig:podk}
\end{figure}
Fig.~\ref{fig:podk} shows that the probability of success in the
protocol with improved fidelity is always less than the probability of
success in the protocol of Bose~\emph{et al.}  Fortunately, there is
only a small difference between the probabilities of both protocols
for the biggest cavity decay rate for which compensation is still
possible, i.e., for $\kappa/2\pi\approx 0.25$ MHz.

So far we have assumed that there is no possibility of photon emission
to modes other than the cavity mode.  Let us now relax this assumption
and investigate the influence of the spontaneous emission decay rate
of the excited state on both teleportation protocols. The spontaneous
atomic emission destroys the quantum information which Alice wants to teleport
to Bob. Such runs of the teleportation protocols are unsuccessful and
should be rejected. However, an event of spontaneous atomic emission
cannot be detected in both schemes and therefore the spontaneous decay
rate of excited state reduces the average fidelities. We can only
suppress this imperfection by taking $\gamma g^2/\Delta^2$,
$\gamma\Omega^2/\Delta^2\ll\kappa$~\cite{chimczak02:_effect}.  The
biggest $\kappa$ for which the compensation is still possible allows
for the choice of $\gamma/2\pi=1$~MHz.  We have generated 20 000
trajectories to compute the average fidelities and the average
probabilities for the parameters
$(\Delta;\Omega;g;\gamma;\kappa)/2\pi=$ $(100;10;10;1;0.265)$ MHz. As
a result we have obtained the average fidelity of $0.972$ and the
average probability of $0.36$ for the Bose~\emph{et al.} protocol and
the average fidelity of $0.978$ and the success rate of $0.31$ for the
modified protocol.  The results indicate that the inability to
distinguish the runs of protocols, in which spontaneous emission
occurs, reduces only slightly the average fidelities when
$\gamma/2\pi=1$~MHz.  The average probabilities of success remain
unchanged.

Other two important imperfections, which we have to take into account,
are a finite detection quantum efficiency and the presence of dark
counts.  It is necessary to include such sources of noise in our
numerical calculations because they are introduced by all real
detectors.  So far we have assumed in our analysis perfect detectors
that are able to register all collected photons and do not produce any
signal in the absence of photons. In practice, this assumption is not
valid. The probability that a single photon reaching the detector is
converted into the measurable signal, which is called the quantum
efficiency and denoted by $\eta$, is less than unity for all real
detectors~\cite{bachorksiazka, carmichaelksiazka}.  Moreover, there
are clicks, for all real detectors, even in the absence of light. They
are called dark counts.  These imperfections lead to lowering the
average fidelity in both teleportation protocols because of randomness
which they introduce to the measurement outcome. There is no way to
distinguish the unsuccessful case of two photon emissions from the
desired case of one photon emission when only one of the two emitted
photons is detected.  It is also not possible to recognize the
unsuccessful case of no emission if one dark count occurs during the
detection stage.  The quantum information that Alice wants to teleport
is destroyed in the unsuccessful cases. If one cannot reject such
cases then the average fidelity is reduced. Therefore it is necessary
to use detectors with very high efficiency $\eta$ and a low enough
dark count rate. As far as we know, the highest detector efficiency
has been reported by Takeuchi~\emph{et al.}~\cite{takeuchi99} and is
equal to $\eta=0.88$.  To study the effect of the detector
inefficiency on the protocols under discussion, we have performed
numerical calculations under the assumption that there are not dark
counts first. We have used the same parameters as previously, i.e.,
$(\Delta;\Omega;g;\gamma;\kappa)/2\pi=$ $(100;10;10;1;0.265)$ MHz and
we have found that both protocols are sensitive to the detector
inefficiency. The average fidelity is reduced to $0.894$ in the
Bose~\emph{et al.}  protocol and to $0.905$ in the modified protocol.
Success rates remain almost unchanged --- $0.353$ in the Bose~\emph{et
  al.} protocol and $0.306$ in the modified protocol. It is obvious
that the reliable teleportation requires detectors efficiency
$\eta=0.88$ or higher. Unfortunately, the dark count rate of the
detector increases roughly exponentially with the
efficiency~\cite{takeuchi99} and is as high as $20$~kHz at the highest
efficiency reported by Takeuchi~\emph{et al.}~\cite{takeuchi99}, i.e.,
$\eta=0.88$. The high efficiency of the detector means also the high
rate of dark counts, which are not good for teleportation. To clarify
the situation, we have also investigated the influence of the dark
count rate on both teleportation protocols. Surprisingly, the protocol
with improved fidelity has appeared to be less sensitive to this
imperfection than the Bose~\emph{et al.} protocol. The average
fidelity in the Bose~\emph{et al.} protocol appeared to be equal to
$0.801$ while the average fidelity in the modified protocol to be
equal to $0.897$, for the parameters $\eta=0.88$ and the dark count
rate $20$~kHz.  The difference between the two protocols is quite
impressive, but it has a simple explanation. In either protocol there
is only one stage when the detection of one photon is expected --- the
detection stage in the Bose~\emph{et al.} protocol and the detection
stage I in the modified protocol. Only in these two stages occurrence
of the dark count can be erroneously accepted as a successful
measurement event because all other stages require no photon detection
to be successful. Thus, one can easily understand why the influence of
the dark counts on both protocols is different by comparing the times
of the two crucial stages --- the time of the detection stage of the
Bose~\emph{et al.} protocol (in our calculations we set
$t_{D}=10\kappa^{-1}$) that is much longer than the time of the
detection stage I ($t_{d}=\pi\delta^{-1}$) of the modified protocol.
This means that there are many more rejected dark counts in the
modified protocol than in the Bose~\emph{et al.}  protocol.  A bigger
number of rejected runs with the dark count events leads to an
increased average fidelity and at the same time to a decreased
success rate.  Therefore, the success rate is reduced more
significantly in the modified protocol ($0.237$) than in the
Bose~\emph{et al.} protocol ($0.331$).

Finally, we generalize our calculations to include losses in the
mirrors and during the propagation. The absorption in the mirrors can
be taken into account by making the replacement
$\kappa=\kappa'+\kappa''$ in the Hamiltonian~(\ref{eq:Hamiltonian0}),
where $\kappa'$ is the decay rate corresponding to the photon
transmission through the mirror and $\kappa''$ is the photon loss rate
due to absorption in the mirrors. The evolution of the system is
conditional, so we need also the collapse operators corresponding to
the absorption of photons in the mirrors. The additional collapse
operators are given by $C_{A}=\sqrt{2\kappa''} a_{A}$ and
$C_{B}=\sqrt{2\kappa''} a_{B}$.  As before, the collapse operators
describing photon detections are given by~(\ref{eq:C}) but with
$\kappa$ replaced by $\kappa'$. So, we now have two extra collapse
operators describing evolution of the system. However, it can be
checked that such evolution can be described without using the extra
collapse operators when we make the replacement
$\kappa=\kappa'+\kappa''$ in the collapse
operators given by equation~(\ref{eq:C}) and multiply
the probability of photon detection  by
$\eta_{a}=(\kappa'/\kappa)$, which is the 
probability that a photon is detected despite the fact that there is
absorption in the mirrors. The probability of detection in the
presence of absorption is then $P'_{D}=\eta_{a}P_{D}$. The presence of
absorption means effectively lower efficiency of the detector.

In the same way, we easily can take into account all photon losses
during the propagation between the cavities and the
detectors~\cite{bose,duan:_effic}. All we need to include such losses
into consideration is to introduce additional efficiency factor
$\eta_{p}$. Multiplying all the factors, we find the overall detection
efficiency $\eta'=\eta_{a}\eta_{p}\eta$. To visualize the effect of
such losses, we have plotted the average fidelity and the average
probability for both protocols as functions of the overall detection
inefficiency, i.e., as functions of $1-\eta'$. In order to make the
average values reliable, we have generated 100 000 trajectories for
each $\eta'$.
\begin{figure}[htbp]
  \centering
  \includegraphics[width=7cm]{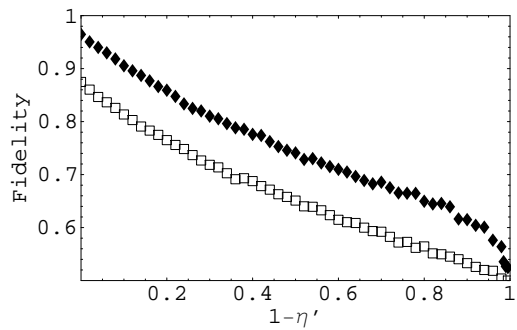}
  \caption{The average fidelity including the effects of photon losses
    as a function of the overall detection inefficiency for
    $(\Delta;\Omega;g;\gamma;\kappa)/2\pi=(100;10;10;1;0.265)$ MHz and
    the dark count rate $20$~kHz.  The diamonds correspond to the new
    protocol and the open squares correspond to the Bose~\emph{et al.}
    protocol.}
  \label{fig:feta}
\end{figure}
From figure~\ref{fig:feta} it is clear that with increasing photon
losses the average fidelity is reduced for both protocols. However,
the advantage of the modified protocol to be less sensitive to the
dark counts and the compensation for the factor $e^{-\kappa t_{A}/2}$
result in the fidelity improvement that is clearly visible for almost
all values of $\eta'$. The difference between both protocols
disappears only for such a small $\eta'$ that most of the trajectories
for which measurement indicates success are unsuccessful cases due to
dark counts. Of course, in such a case the final state of Bob's atom
is random and the average fidelity is $0.5$.

\begin{figure}[htbp]
  \centering
  \includegraphics[width=7cm]{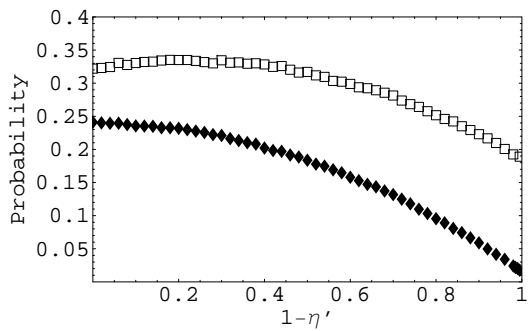}
  \caption{Average probabilities that measurement indicates success
    for the modified protocol (diamonds) and for the Bose~\emph{et
      al.} protocol (open squares) as functions of the overall
    detection inefficiency.  The parameter regime is
    $(\Delta;\Omega;g;\gamma;\kappa)/2\pi=(100;10;10;1;0.265)$ MHz,
    the dark count rate is $20$~kHz.}
  \label{fig:peta}
\end{figure}
From Fig.~\ref{fig:peta} it is visible that higher fidelity can be
achieved by accepting lower success rates.  The average probability of
success in the modified protocol is always less than the average
success rate in the Bose~\emph{et al.} protocol. This is the price we
have to pay for higher fidelity.

\section{Conclusions}
We have presented the teleportation protocol for the device proposed
by Bose~\emph{et al.}  that improves the fidelity of teleported
state. The improvement is obtained by compensating for the factor
$e^{-\kappa t_{A}/2}$ which appears in the teleportation protocols. We
have shown that this compensation makes it possible to stabilize the
fidelity at a high level despite the increase in the cavity decay
rate. The fidelity is stabilized until $\kappa/2\pi\approx 0.25$ MHz.
This means that the high fidelity teleportation can be performed for
the values of the cavity decay rates over $25$ times larger than the
values assumed by Bose~\emph{et al.}. The price we have to pay for
more realistic values of the cavity decay rates is that
we have to accept lower success rates. We have also shown that the
modified protocol is less sensitive to the dark counts of detectors
than the original protocol of Bose~\emph{et al.}

\section*{ACKNOWLEDGMENTS}
This work was supported by the Polish Ministry of Science and Higher
Education under Grant No.~1~P03B~064~28.


\end{document}